\documentclass{camera}

\usepackage{nicefrac}
\usepackage{graphicx}

\begin{document}

\title{NEUTRINOS FROM SUPERNOVA REMNANTS
AFTER THE FIRST H.E.S.S.\ OBSERVATIONS}

\author{Francesco Vissani}
\organization{INFN, LNGS Theory Group, Assergi (AQ) Italia}

\maketitle

{\centerline{\bf Abstract}
\small 
\vskip1mm
\noindent 
We provide elements for a discussion 
of the expected $\nu$ signal from 
Supernova Remnants (SNR) in the Milky Way.
After recalling why SNR are interesting and 
certain remarkable achievements of H.E.S.S., we 
describe a simple and straightforward 
method to evaluate the $\nu$ fluxes from
$\gamma$-ray data. 
For an ideal detector, 
we get a flux of 5  thoroughgoing muons per km$^2$ 
per year from RX J1713.7-3946 
in ANTARES location and above 
$E_{th}=50$ GeV; similar calculations for Vela Jr 
show that the number of events, to be evaluated precisely 
after the next detailed observations of H.E.S.S.,
are larger.
We comment on the role of neutrino oscillations.}

\section{The cosmic ray/SNR connection}

Supernovae are suspected to be the cosmic ray (CR)
accelerators since '34 (Baade \& Zwicky \cite{bz}).
30 years later, Ginzburg \& Syrovatsky \cite{gs} remarked that if 
10 \% or so of the SNR kinetic energy ${\cal E}_{SN}\approx 
10^{51}$ erg (=1 foe, also known as 1 bethe) goes in CR, 
the Milky Way losses are compensated: 
$$
\frac{V_{CR}\ \rho_{CR}}{\tau_{CR}}\approx 0.1
 \times \frac{{\cal E}_{SN}}{\tau_{SN}}
$$
where $V_{CR}=\pi R^2 H$ ($R=15$ kpc, $H=5$ kpc),  $\tau_{CR}=10^7$
yr and $\tau_{SN}=30$ yr.
The `diffusive shock wave acceleration' 
mechanism based on Fermi ideas~\cite{fer}
is being developed to explain CR acceleration in SNR 
(lecture of Blasi); 
CR accelerate in an expanding shock wave  
of size $R=u\ t$ ($u\sim$ 5,000 km/s), mostly 
active in the first 1,000  yr, determined by 
$M_{ejecta}\sim$ $\nicefrac{4\pi}{3}\  R^3\ n_{ISM}$.

Hillas selected a list of open (connected?) questions \cite{hill}:

\noindent
$\bullet$
How to ``inject'' electrons? $\langle$ {\em diffusive shock accel.\ is
incomplete?} $\rangle$

\noindent  $\bullet$
Why isotropy? How $\Gamma=2.1\to 2.7$? $\langle$ 
{\em propagation/reacceleration?} $\rangle$

\noindent  $\bullet$
$E_{\rm max}$? $\langle$ {\em 
 $R\sim D_{Bohm}/u$ limit, countered by Bell \& Lucek \cite{bl}}
$\rangle$

\noindent  $\bullet$ Too few point sources of VHE $\gamma$? 
$\langle$ {\em something lacking?} $\rangle$

\noindent  $\bullet$ How to firmly exclude a leptonic origin? 
$\langle$ {\em TeV $\nu$ are needed?} $\rangle$

\noindent 
We will be mostly concerned with the last 
questions (the last one is not in Hillas's list, but perhaps he 
would have included it, if giving this talk).

\subsection{The landscape after the first H.E.S.S.\ results}
For a few young shell-type SNR observed in VHE $\gamma$, the  
``hadronic'' hypothesis seems plausible; more crucial  tests
will be possible with future observations by H.E.S.S.\ 
(VERITAS, MAGIC) and other instruments. 
The closest SNR's, whose properties we recall here,
are of particular interest:   
\vskip2mm
\centerline{\small
\begin{tabular}{l|c|c|c|c|c}
Name & TeV $\gamma$ observ.\ & decl.\ $\delta$ & distance & size & age \\ \hline
Vela Jr & $<10$ TeV (HESS) & $-46^\circ 22'$ & 0.2 kpc & $2^\circ$ & 680 yr\\  
RXJ1713... & $<40$ TeV (HESS) & $-39^\circ 46'$ & 1 kpc & $1^\circ$ & 1,600 yr\\  
SN 1006 & no(t yet?)  & $-41^\circ 53'$ & 2 kpc & $36'$ & 1,000 yr\\  
Cas A & HEGRA (maybe) & $58^\circ 08'$ & 3 kpc & $6'$ & 320 yr\\  
\end{tabular}}
\vskip2mm
\noindent 
(note however that the  ``distance'' and ``age'' are not reliably 
determined, see e.g.~\cite{chandra}). 
For the first two the velocity of expansion
is as expected, for the other it is a bit larger.   
RX J1713.7-39346 and Vela Jr are the most intense shell-type SNR in 
the TeV $\gamma$ sky; thus, it is particularly important to discuss the
expected neutrino fluxes from them.

The best known $\gamma$ spectrum is the one of RX J1713.7-3946.
H.E.S.S.~\cite{hsrx} showed that the spectrum 
of RX J1713.7-3946 deviates from a
power law distribution above $\sim 10$ TeV. Assuming
that the $\gamma$-rays come from CR interactions with the 
environment (are of hadronic origin), the cut in the CR (proton) spectrum
should be around 150 TeV, in agreement with 
naive expectations from diffusive 
shock acceleration models. 
Specific models of this SNR have been proposed:
Malkov, Diamond, Sagdeev '05 \cite{mod1} suggest that the nearby 
molecular cloud has a main role for CR interactions whereas
Berezhko \& V\"olk '06 \cite{mod2} fit H.E.S.S.\ observations
starting from the opposite view; in both models the
hadronic contribution is sizable or dominant.
In short, H.E.S.S.\ observations renewed the interest in the 
CR/SNR connection and perhaps, figured out the most intense sources
of VHE $\gamma$ rays. 

Two web links that 
can help to keep information updated are:\\
$\star$ {\sf H.E.S.S.\ Source Catal.}, {\tt www.mpi-hd.mpg.de/hfm/HESS/} 
[W.~Hofmann];\\[-.2ex]
$\star$ {\sf Catalogue of SNR}, {\tt www.mrao.cam.ac.uk/surveys/snrs/} 
[D.~Green].\\
See also 
the review of H.J.~V\"olk \cite{vvv}
and the lecture of Wei Cui.

\section{TeV neutrinos from SNR}
{\em Motivated 
by the (shell-type, young) SNR / CR connection
and by the existing plans for large neutrino telescopes, 
we calculate  the flux of TeV neutrinos from the SNR with known
VHE $\gamma$-ray spectrum.}
Indeed, during CR acceleration the 
SNR  are transparent to their $\gamma$ radiation. 
Thus, 
we can convert the measured $\gamma$ ray flux (from $\pi^0$ and $\eta$)
into an expectation for the neutrino flux
(from $\pi^\pm$ and $K^\pm$)
under the hypothesis that the radiation is of hadronic origin.
We begin by discussing flavor oscillations, describe the $\gamma/\nu$
connection and estimate the rate of events in (km$^2$ class, ideal)
neutrino telescopes.
\subsection{Oscillations}
The flux of neutrinos--from meson decays--are modified by 
the oscillations: 
$$
F_{\nu_\mu}=F_{\nu_\mu}^0 P_{\mu\mu} +
F_{\nu_e}^0 P_{e\mu} 
$$
The oscillation probabilities 
take the simplest form, Gribov-Pontecorvo's~\cite{gp}
(namely, the one that applies for low energy solar neutrinos):
$$
P_{\ell \ell'}=\sum_{i=1}^3 |U_{\ell i}^2|\ |U_{\ell' i}^2| \ \ \
\mbox{ with }\ell,\ell'=e,\mu,\tau
$$
There is no MSW effect \cite{msw}, 
for matter term is negligible close to the SNR and too large in
the Earth. With central values of the mixing elements $U_{\ell i}$ we get 
$P_{\mu \mu}\sim 0.4$ and $P_{e \mu}\sim 0.2$; that is,  
$\nicefrac{1}{2}$ of the original $\nu_\mu$ and $\bar{\nu}_\mu$ 
fluxes reach the detector.

We performed also a detailed (or sophisticated) analysis
$$
{\cal L}(P_{\mu\mu})\propto {\rm max}\! \left[
e^{-\frac{(P_{\mu\mu}-P_{\mu\mu}(\theta))^2}{2\sigma^2}}
\times 
{\cal L}_{osc.}(\theta) 
 \right]
\mbox{ with }\sigma\to 0
$$
where $\theta$ are the measured parameters taken  
from \cite{sv} that we marginalize away 
by maximizing the result.
We get 
$P_{\mu \mu} = 0.39\pm 0.05$ and $P_{e \mu} = 0.22\mp 0.05$ where
most of the error (0.04) is due to $\theta_{23}$.

To understand the uncertainty budget one can use 
an expansion in the small parameters \cite{cv}:
$$
\begin{array}{l}
P_{\mu\mu}\simeq \nicefrac{1}{2}-\nicefrac{x}{8}-y \mbox{ and }
P_{e\mu}\simeq \nicefrac{x}{4}+y, \\[1ex]
\ \ \ \ \ \ \ \ \ \ \mbox{where }\left\{ \begin{array}{l} x=\sin^2 2\theta_{12},\\
y=\cos 2\theta_{23}\ \nicefrac{x}{4} +
\theta_{13}\ \cos\delta_{\rm CP}\ \nicefrac{\sqrt{x(1-x)}}{2} .
\end{array} \right.
\end{array}
$$
In view of astrophysical uncertainties and small counting rates
we believe that the uncertainties in the oscillation parameters do not have 
  an important role for the discussion of SNR $\nu$ and 
  presumably even for other cosmic sources.\footnote{
We comment on a proposal 
\cite{mic} to study $\theta_{13}$ and $\delta_{\rm CP}$ 
with a source of $\bar{\nu}_e$ 
\cite{anc,ter} \label{michele} using the $\bar{\nu}$ flux ratio
${F_\mu}/{(F_e+F_\tau)} ={P_{e\mu}}/{(1-P_{e\mu})}$.
Since the shift due to $\delta_{\rm CP}$  and
$\delta\theta_{13}$ around $0^\circ$ 
is $\delta P_{e\mu}=0.02$, first we should know 
the impact of $\theta_{23}$.
If this were negligible, the number of 
$e+\tau$ signal events $N$ should 
obey $\delta N/N \le 10$~\%. 
This needs 60 years of an ideal detector 
and a small systematic $\delta b/b \le 1$~\%,
if we use the event rates  
per year in a km$^2$ area of~\cite{mic}
$s=16$ (signal) and $b=145$ (background): $10^3$ 
signal events over $10^4$ events.
We get instead $s=s_e+s_\tau=1.3$
($s_e=N_t \int_{1\ \rm{TeV}}^\infty \! \sigma(E)\ F_e(E)\ dE$;
targets: $N_t=4.5\ 10^{38}$ nucleons; 
flux from \cite{anc}, fig.1: 
$7\ 10^{6}$ $({E}/{\mbox{TeV}})^{-3.1}$ 
$ 1/\mbox{TeV yr km}^2$ with $P_{ee}\approx 0.6$;
cross section: $4\ 10^{-36} (E/\mbox{TeV})^{0.87}\mbox{ cm}^2$).
Even assuming the source is real, with $s=1.3$ 
the question is 
whether a signal can be seen.} 

\subsection{The connection between $\gamma$ and $\nu$}
For RX J1713.7-3946 there are various calculations in the literature:
\begin{enumerate}
\item Alvarez-Mu\~niz \& Halzen '02 \cite{amh} inspired by CANGAROO
first observations 
use $F_\gamma\propto E^{-2}$ and obtain
$F_{\nu_\mu}=F_{\nu_\mu}^0 \propto F_\gamma $ by
$$
\int_{E_p^{\rm min} /12}^{E_p^{\rm max} /12} 
\!\!\!\! \!\!\!\! \!\!\! \!\!\!\! \!\!\!\! dE_\nu\ 
E_\nu F_\nu(E_\nu)=
\int_{E_p^{\rm min} /6}^{E_p^{\rm max} /6} 
\!\!\!\! \!\!\!\! \!\!\!\! \!\!\!\! \! dE_\gamma\ 
E_\gamma F_\gamma(E_\gamma)
$$
\item Costantini \& V '04 \cite{cv} use 
$F_\gamma\propto E^{-2.2}$ as extrapolated from early H.E.S.S.\ 
below 10 TeV results
and adopt standard techniques (see e.g., \cite{gais,next})
$$
F_\gamma=\frac{\Delta X}{\lambda_p}\ \frac{2 Z_{p\pi^0}}{\Gamma} F_p\ 
\mbox{ and similarly for } F_\nu
$$
\end{enumerate}
Both methods, however, are tailored for power law spectra; and we
know from the only detailed observation that we have 
(RX J1713.7-3946) that this is not a good
approximation. Thus we are lead to recalculate  the neutrino fluxes.
In principle, one could de-convolute the CR
flux from the $\gamma$-ray flux and then obtain the 
neutrino flux, as described in the lecture of K.-H.~Kampert. In our case, 
when the atmosphere (the target for CR) is much thinner--it 
is transparent--a much simpler and direct 
approach \cite{vis06} is possible based on 
the classical techniques of Lipari '88 \cite{lipas}.
In fact, from the integral expression for VHE $\gamma$-rays
$F_\gamma(E)=\int_E^\infty dE'\ 2\ F_{\pi^0}(E')/E'$
we find immediately:
\begin{equation}
F_{\pi^0}(E)=-\frac{E}{2}\ \frac{dF_\gamma}{dE}
\label{1}
\end{equation}  
Due to the approximate isospin-invariant distribution
of pions, $F_\pi \equiv F_{\pi^0} \approx F_{\pi^+}\approx F_{\pi^-}$, 
we find for the neutrinos from $\pi^+\to \mu^+ \nu_\mu$:
\begin{equation}
F_{\nu_\mu}(E) =\int_{E/(1-r)}^\infty \frac{dE'}{1-r}\ \frac{F_\pi(E')}{E'}
=\frac{F_\gamma(E/(1-r))}{2(1-r)}
\label{nu1}
\end{equation}
where $r=(m_\mu/m_\pi)^2$. The neutrinos $\nu=\bar{\nu}_\mu, \nu_e$
from $\mu^+$ decay are:
\begin{equation}
F_{\nu}(E_\nu)= 
\int^1_0 \frac{dy}{y}\ F_\mu(E_\mu)\ [ g_0(y)- \bar{P_\mu}(E_\mu)\  g_1(y) ]
\mbox{ where }E_\mu = \frac{E_\nu}{y}
\label{nu2}
\end{equation}
$g_i$ are polynomials, 
$F_\mu$ and $\overline{P}_\mu$ 
(polarization averaged over $\pi$ distribution)
are also known. The contributions of 
semileptonic $K^\pm$ decays to the $\nu$'s and the one  
of $\eta\to \gamma \gamma$ to $\gamma$-rays 
can be described with a contribution proportional to the 
pionic one; these
two correct
the formulae in opposite directions so that {\em effectively} one 
should just add a contribution of kaons of the order of 10
percent. For details, see~\cite{vis06}.

\begin{figure}[t]
\caption{\em $\nu_\mu$ spectra, corresponding to two fits of the 
H.E.S.S.\ VHE $\gamma$-rays from RX J1713.7-3946 \cite{hsrx}: 
a broken power law and a power law with 
exponential
cutoff; the third fit--a curved power law--is 
incompatible with a hadronic origin,
for it increases before {\rm 40 GeV.}
The $\bar{\nu}_\mu$ flux is the same in our approximation.
\label{fig1}}
\vskip-4mm
\centerline{
\includegraphics[width=0.4\textwidth,angle=270]{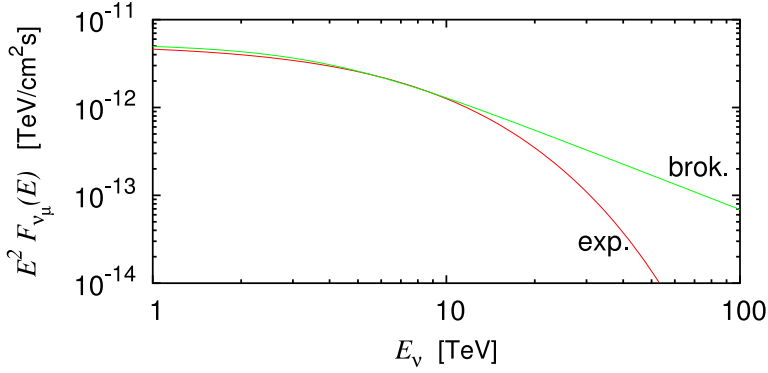}}
\end{figure}

\subsection{A case study: RX J1713.7-3946}
The neutrino flux, evaluated as described above, 
is shown in 
figure~\ref{fig1} and table~\ref{tab1}. For the purpose of illustration,   
we use it to calculate the number of muon events 
$N_{\mu}+N_{\bar\mu}$ for an 
{\em ideal detector}\footnote{By definition, 
`ideal detector' means that all muons above threshold are detected:
$\epsilon(E_\mu)=1$ for $E_\mu>E_{th}$. 
Of course in a 
real detector (where the geometrical area $A$ changes with the time 
of observation
$t$ and, above all,  
the efficiency $\epsilon$ increases with the energy)
the impact of the cut in the spectrum 
is expected to be much more important.} 
using: 
$$
N_{\mu}=\displaystyle 
A\cdot T\cdot f_{liv}\cdot 
\int_{E_{th}}^{\infty} dE_\nu\; 
F_{\nu_\mu}(E_\nu) Y_{\mu}(E_\nu,E_{th}) 
( 1-\overline{a}_{\nu_\mu}(E_\nu) )
$$
where $E_\nu$ is the neutrino energy before the 
interaction point and the various quantities in the previous formula 
are defined as follows:\\ 
$\bullet$  $A=1$ km$^2$ and $T=1$ solar year.\\
$\bullet$  The source is below ANTARES horizon (=visible) 
for $f_{liv}=78$ \%.\\
$\bullet$  The threshold for muon detection is $E_{th}=50$ GeV (as low
as possible).\\
$\bullet$  The muon range (that goes 
in the yield $Y_\mu$) is calculated for water.\\
$\bullet$  The neutrino absorption coefficient $a_{\nu_\mu}$, 
averaged over the daily location of the source, is calculated for 
standard rock.

We find that the total number of events 
does not depend crucially on the extrapolation:
$$
N_{\mu}+N_{\bar{\mu}}=\left\{
\begin{array}{ll}
\mbox{4.8 per km$^2$ per year} &\mbox{ [exponential cutoff] }\\
\mbox{5.4 per km$^2$ per year}&\mbox{ [broken power law]}
\end{array}
\right.
$$
This can be compared with the 9 events in  \cite{cv}
(power law extending till 1 PeV) and the 40 events in \cite{amh} 
(oscillations, livetime and absorption ignored).

{\begin{table}[b]
{\begin{center}
\footnotesize
\begin{tabular}{c|ccccccccc} 
\hline \hline
$E$& 0.1 & 0.2 & 0.4 & 0.6 & 0.8 & 1 & 2 & 4 & 6  \\ \hline
exp.& 
5.1e2 & 1.3e2 & 3.1e1 & 1.4e1 & 7.4e1 & 4.6e1 & 1.0e0 & 1.9e{{-1}} & 6.1e{{-2}}
\\ 
brok.& 
5.9e2 & 1.4e2 & 3.4e1 & 1.4e1 & 7.9e0 & 5.0e0 & 1.1e0 & 1.9e{{-1}} & 6.1e{{-2}}  \\ \hline \hline
$E$ & 8 & 10 & 20 & 40 & 60 & 80 & 100 & 200 & 300  \\ \hline
exp. & 
2.6e{{-2}} & 1.3e{{-2}} & 8.7e{{-4}} & 2.4e{{-5}} & 1.4e{{-6}} & 1.2e{{-7}} & 1.3e{{-8}} & 4.4e{{-13}} & 3.0e{{-17}}  \\ 
brok. & 
2.6e{{-2}} & 1.3e{{-2}} & 1.4e{{-3}} & 1.4e{{-4}} & 3.7e{{-5}} & 1.4e{{-5}} & 6.9e{{-6}} & 7.0e{{-7}} & 1.8e{{-7}} \\ \hline \hline
\end{tabular}
\end{center}}
\vskip-4mm
\caption{\em Differential $\nu_\mu$ flux from RX J1713.7-3946 
for selected energies ($1^{st}$ and $4^{th}$ lines).
Units: {\small \rm TeV} for the energy,  
{\small \rm 1e-12/}$\mbox{\rm TeV s cm}^2$ for the flux.
\label{tab1}}
\end{table}}

\section{Summary and perspectives}
We discussed the expected 
neutrino flux from $\gamma$-transparent 
accelerators of cosmic rays, and emphasized 
the case of young SNR.
We showed that for RX J1713.7-3946 
(the best known SNR in TeV sky, thanks to H.E.S.S.) 
the expectations are stable: 
$\sim 5$ events per km$^2$ per year in an ideal detector.
The median neutrino energy is $3$ TeV. 
Since the detected $\mu$ are softer than the $\mu$ 
in the production point (that in turn is softer 
than the impinging $\nu$) several events will fall 
in an energy region where the atmospheric background and 
the role of imperfect detection efficiency are important:
see the lectures of Lipari and Lucarelli.
Thus, we believe that it would be desirable to have
a detailed discussion of the characteristics of a detector 
that aims to see a $\nu$ signal from RX J1713.7-3946. 

\vskip2mm

Sometimes soon H.E.S.S.\ should tell us more on another 
intense VHE $\gamma$-ray source, RX J0852.0-4652 (Vela Jr).
This SNR
has $F_\gamma=6.5$ million $\gamma$-rays 
$E^{-2.1}/(\rm TeV\ km^2\ yr)$ below $10$ TeV.
If the exponential cutoff is at $E_{\rm \gamma cut}=50$ $(150)$ TeV
(and again, if these $\gamma$ rays are of hadronic origin)
we get $N_\mu + N_{\bar\mu}=10$ $(14)/(\rm km^2 yr)$ in an ideal detector, 
with a significantly higher energy~\cite{vis06}.

\subsection*{Acknowledgments}
I gratefully thank 
F.~Aharonian,
B.~Aschenbach,
R.~Aloisio,
J.~Beacom,
V.~Berezinsky, 
A.~Butkevich,
M.~Cirelli,
P.~Desiati,
C.~Distefano,
S.~Dugad,
W.~Fulgione,
P.L.~Ghia, 
D.~Grasso,
W.~Hofmann,
T.~Montaruli,
G.~Navarra,
I.~Sokalsky,
A.~Strumia,
R.~Thorne,
Y.~Uchi\-yama
and especially 
M.L.~Costantini and 
P.~Lipari 
for useful discussions and help.

\appendix\section{Details of signal evaluation}
Following \cite{cv} and \cite{vis06},
we describe here two specific elements used for the 
evaluation of the signal and comment on them.
\subsection{Muon background}
It should be possible 
to increase the acceptance by designing an 
angular cut that depends 
on the energy and on the time of the event. 
Yet, here we follow the simplest prescription to exclude 
cosmic muon contamination: 
accept only the events below the horizon of the detector. 
The angle $\theta$ 
between an astronomical object and the vertical of the detector
is
$$
\cos\theta(t)=(\ \cos\delta,0,\sin\delta\ )\cdot 
(\ \cos\phi\cos(\pi t/\tau),\cos\phi\sin(\pi t/\tau),\sin\phi\ )
$$
where 
{\em 1)} $\delta$ is the declination of the object (for the two 
sources discussed in the text: 
RX J1713.7-3946 $=-39^\circ 46'$,
RX J0852.0-4652 / Vela Jr $=-46^\circ 52'$);
{\em 2)} $\phi$ is the latitude of the detector 
(for the detectors in the Northern hemisphere:
Baikal $=51^\circ 50'$,
ANTARES $=42^\circ50'$,
NEMO $=36^\circ 30'$,
NESTOR $=37^\circ 33'$); 
{\em 3)} $2\times \tau= 23^h 56^m 4^s$ is the duration
of the sidereal day; 
{\em 4)} $t$ is the time measured from the point when 
the object is at the apex. If 
$\cos\theta(t=0)<0$,  
the object remains always below the horizon; if the converse 
happens, but $\cos\theta(t=\tau)<0$, 
the object is observable for a fraction of the time 
$1-\tau_0/\tau$, determined by the
condition $\cos\theta(\tau_0)=0$. 

\subsection{Earth absorption\label{prevas}}
The Earth absorption of the neutrino depends on the time $t$ 
through the average column density in the Earth and it 
is mostly due to CC interactions. NC interactions are 
$\sim \nicefrac{1}{3}$ smaller and, furthermore, 
cannot remove a neutrino, but  
only degrade its energy.
A detailed evaluation  of this effect is 
numerically demanding \cite{terry}, 
however it supports the expectations from 
a simple heuristic argument inspired to the formalism 
of `scaling' (proper of 
strong interaction): NC interactions increase the 
absorption, but only by few~\%. Thus in a first approximation
NC effect can be neglected.

\subsection{Comments}
In the light of the new H.E.S.S.\ results, it is useful to note that:
\begin{enumerate}
\item Accepting 
events from a few degree above the horizon (that is, setting the cut
to a certain value $\cos\theta_0>0$) can lead to 
significant increase of the events, especially for 
RX J1713.7-3946 \cite{distefy}.
E.g., for ANTARES the fraction of time 
becomes $f_{liv}=88$ \% just adding 
5 degrees above the horizon.\footnote{For
a cosmic muon that reaches the depth of 1 km under the surface,
$5^\circ$ means more than 12 km of water but recall that a muon
can change its direction by scattering.}
\item The absorption coefficient obtained by the approximate
procedure mentioned in sect.\ \ref{prevas},  
using a power law distribution $E^{-2.2}$
and  averaged over the fraction of the time when $\cos\theta(t)<0$
is very similar for RX J1713.7-3946 and for 
Vela Jr (slightly weaker in the second case).
\item If the neutrinos are confined 
to be of relatively low energy, as it is the case of 
RX J1713.7-3946, the effect of Earth absorption is small.
\end{enumerate}

{

}

\section*{Discussion}
\paragraph{AURELIO GRILLO:}
1)~How reliable is the $\gamma$-transparency hypothesis?\\
2)~What is the relative weight of oscillations and correct spectrum?
\paragraph{FRANCESCO VISSANI:}
1)~For a SNR as RX J1713.7-3946 it is possible to
check that the matter is very diffuse either distributing several 
solar masses in the wide volume, or more directly using
$X$-ray data, so that the absorption of $\gamma$-rays is
negligible. E.g., for the molecular cloud: 
100 $M_\odot$ at 1 kpc and a size of $15'$ means 
a column density of about 
$2\ 10^{20}$ protons/cm$^2$
(10 protons per cm$^3$).
For other sources as $\mu$QSO (namely, 
stellar black holes
or neutron stars with jets) this does not 
apply and a relatively weak $\gamma$-radiation could 
be compatible with intense $\nu$ 
fluxes \cite{dist}.\footnote{It was suggested 
that $\mu^\pm$ (but not $\pi^\pm$) are absorbed in $\mu$QSO. This
leads to a peculiar flavor neutrino ratio, as for
$\bar{\nu}$\ from hypothetical neutron sources (see 
footnote \ref{michele}).
Of course, before testing flavor ratios we should  
be sure that we can at least observe  muon neutrinos above TeV
and the most cautious attitude would suggest 
to perform all possible tests with conventional means first.}\\
2)~Again for RX J1713.7-3946, each of the two effects  
reduce the signal in an ideal detector by a
factor of $\nicefrac{1}{2}$. 
However oscillations are universal whereas 
deviations from power-law depend on the individual object. 
We hope this is less pronounced for Vela Jr, and wait for the response of 
H.E.S.S.

\vskip.1cm 

\paragraph{GIANNI NAVARRA:} Is it possible to ``extrapolate'' these
calculations
in order to obtain a neutrino luminosity from the galactic disk, in
the hypothesis that all CR are produced by SNR?
\paragraph{FRANCESCO VISSANI:} 
Strictly speaking the results in eqs.\ref{1},\ref{nu1},\ref{nu2} 
are just a tool to assess 
an upper bound to $\nu$ from $\gamma$-transparent sources
(or if you like it more, a lower bound to $\nu$ 
from VHE $\gamma$ ray sources of hadronic origin)
and I can only hope that they will be useful for tasks as
the one you propose.\footnote{Just after Vulcano 2006 
several papers appeared addressing this type of problem
\cite{lipp,jb,ahah2}, mostly using assumptions 
on the CR flux at least for intermediate
steps. More information and discussion 
in two subsequent conferences \cite{cris} and \cite{tev}.} 
One ingredient for the calculation seems to be the following: 
if we have a new SN each 30 years 
and if they are mostly active for some thousand  years,
we have 40-80 SNR that inject CR effectively 
in each moment. 
These are most likely located in 
the spiral arms of the Milky Way,
but we can get
a rough idea of their distribution assuming that their
density is $r \exp(-r/r_0)$, with $r_0\sim 3$ kpc 
\cite{ifae}. In this way we can find   
the expected distance of the closest one, 
their average distance, etc. These results could/should 
be cross-checked with SNR databases. 
An obstruction could arise in the description of an `average' SNR:
at different ages the intensity and the distribution 
of $\gamma$-rays could change.
In fact, I suspect that the SNR  that we 
are beginning in the $\gamma$-rays sky
are peculiar objects, as those associated
with molecular clouds, and/or the brightest/closest ones.

\vskip.1cm 

\paragraph{BERND ASCHENBACH:} Just a comment: Even if it should turn out
that the TeV spectrum in RX J1713.7-3946 is dominated by hadronic processes, 
it is also very useful to compute the contribution of leptonic
processes, since it should be there if adiabatic diffusive acceleration
(involving protons and electrons) at the SNR shock front is
universal. The question is at which level.
\paragraph{FRANCESCO VISSANI:} Certainly this task is 
of paramount importance.  Despite the difficulties, 
it should profit from the accumulation of 
new observations, it can be approached in concrete models of 
CR acceleration (such as \cite{mod1} and \cite{mod2})
and it is presented in a moderately optimistic perspective in a
recent authoritative review work \cite{h2}.
\vskip2mm
But I would like to use this occasion to 
recall that, beside the contribution from leptons to $\gamma$
rays spectrum, there is also a hadronic contribution to the spectrum of
hard $X$-rays. This originates from the non-thermal 
population of electrons produced in muon 
decays \cite{alo} (incidentally, no new formula is needed 
to describe  the flux of electrons or positrons
for this is just the same as the one of 
$\bar\nu_e$ or $\nu_e$, see eq.~\ref{nu2}) that could provide us
with a further handle to disentangle the components 
of leptonic and hadronic origin.

\end{document}